\documentclass[aps,prb,superscriptaddress,twocolumn,showpacs,preprintnumbers,amsmath,amssymb]{revtex4}
\usepackage[dvips]{graphicx,color}% Include figure files
\usepackage{dcolumn}% Align table columns on decimal point
\usepackage{bm}% bold math
\usepackage{ulem}
\usepackage{soul}
\begin{document}

\title{The spin Hall effect as a probe of nonlinear spin fluctuations}
\author{D. H. Wei}
\affiliation{Institute for Solid State Physics, University of Tokyo, 5-1-5 Kashiwa-no-ha, Kashiwa, Chiba 277-8581, Japan}
\author{Y. Niimi}
\email{niimi@issp.u-tokyo.ac.jp}
\affiliation{Institute for Solid State Physics, University of Tokyo, 5-1-5 Kashiwa-no-ha, Kashiwa, Chiba 277-8581, Japan}
\author{B. Gu}
\affiliation{Advanced Science Research Center, Japan Atomic Energy Agency, Tokai 319-1195, Japan}
\affiliation{CREST, Japan Science and Technology Agency, Sanbancho, Tokyo 102-0075, Japan}
\author{T. Ziman}
\affiliation{Institut Laue Langevin, Bo\^\i te Postale 156, F-38042 Grenoble Cedex 9, France}
\affiliation{Laboratoire de Physique et Mod\'{e}lisation des Milieux Condens\'{e}s (UMR 5493), Universit\'{e} Joseph Fourier and CNRS, Maison des Magist\`{e}res, BP 166, 38042 Grenoble, France}

\author{S. Maekawa}
\affiliation{Advanced Science Research Center, Japan Atomic Energy Agency, Tokai 319-1195, Japan}
\affiliation{CREST, Japan Science and Technology Agency, Sanbancho, Tokyo 102-0075, Japan}

\author{Y. Otani}
\affiliation{Institute for Solid State Physics, University of Tokyo, 5-1-5 Kashiwa-no-ha, Kashiwa, Chiba 277-8581, Japan}
\affiliation{RIKEN-ASI, 2-1 Hirosawa, Wako, Saitama 351-0198, Japan}

\date{September 11, 2012}

\begin{abstract}
The spin Hall effect and its inverse play key roles in spintronic devices since they allow conversion of charge currents to and from spin currents. The conversion efficiency strongly depends on material details, such as the electronic band structure and the nature of impurities. Here we show an anomaly in the inverse spin Hall effect in weak ferromagnetic NiPd alloys near their Curie temperatures with a shape independent of material details, such as Ni concentrations. By extending Kondo's model for the anomalous Hall effect, we explain the observed anomaly as originating from the second-order nonlinear spin fluctuation of Ni moments. This brings to light an essential symmetry difference between the spin Hall effect and the anomalous Hall effect which reflects the first order nonlinear fluctuations of local moments. Our finding opens up a new application of the spin Hall effect, by which a minuscule magnetic moment can be detected.
\end{abstract}

%\pacs{72.25.Ba, 72.25.Mk, 75.70.Cn, 75.75.-c}% PACS, the Physics and Astronomy

\maketitle

Spintronics has provided an impetus for the intensive development of techniques to generate or detect spin currents, such as nonlocal spin injection~\cite{1}, spin pumping~\cite{2} and spin light emission devices~\cite{3}. The spin Hall effect (SHE)~\cite{4}, which converts a charge current to a spin current, and its inverse have been widely explored in semiconductors~\cite{3,4,5} and nonmagnetic metals~\cite{2,6,7} by using such techniques. While most of the work has concentrated on the phenomenon and mechanism of the SHE itself, Uchida \textit{et al}. have used the SHE as a detector of a spin dependent chemical potential arising from the spin Seebeck effect~\cite{8,9}. Further applications of the SHE are, however, still lacking. Recently the doping of gold with individual iron impurities was considered as a mechanism to enhance the SHE through spin fluctuations of the iron moments~\cite{10}. This raises the question whether the SHE might be sensitive to collective fluctuations, in particular close to a magnetic phase transition where the fluctuations are strong. As for the anomalous Hall effect (AHE) in a pure ferromagnetic metal such as Fe~\cite{11} and Ni~\cite{11,12}, a sharp peak of the Hall resistivity ($\rho_{\rm H}$) appears only below the Curie temperature ($T_{\rm C}$). This can be understood as an effect of fluctuations of local magnetic moments~\cite{13}. In this report we shall demonstrate experimentally, and provide a theoretical argument, that the inverse SHE (ISHE) also shows an anomaly, but that it is substantially different from that of the AHE. Figure 1 shows schematic images of the AHE and the ISHE of a weak ferromagnetic material near $T_{\rm C}$. In the AHE, a charge current is injected into the ferromagnetic material, and the difference of the skew scattering probabilities for spin-up and spin-down electrons results in a Hall voltage. In the ISHE, on the other hand, a spin current is injected and both spin-up and spin-down electrons are deflected to the same side, which also results in a Hall voltage. Contrary to the AHE, however, we observe a dip and peak in the ISHE for weak ferromagnetic NiPd alloys below and above $T_{\rm C}$, respectively. Since the quantity of the involved Ni ions are very small, the present system could be used for sensitive magnetometry. 

\begin{figure}
\begin{center}
\includegraphics[width=6cm]{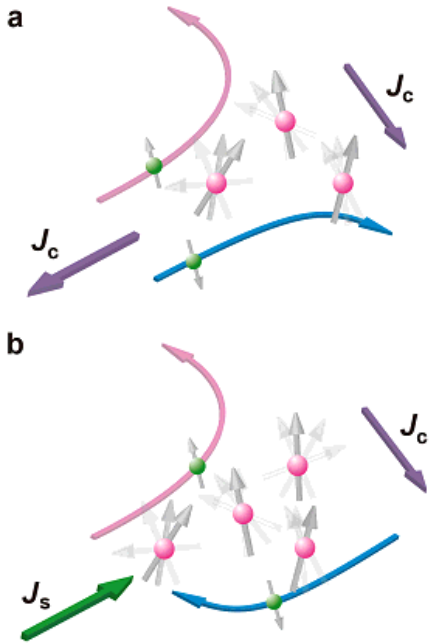}
\caption{\textbf{Principles of the anomalous Hall and inverse spin Hall effects near $T_{\rm C}$ in a weakly ferromagnetic metal.} \textbf{a}, AHE (incident charge current $J_{\rm C}$) and \textbf{b}, ISHE (incident spin current $J_{\rm S}$). The green and pink spheres represent the conduction electrons (i.e., $-|e|$) and localized ones, and the grey arrows show the directions of spins. The fluctuation of the localized spins near $T_{\rm C}$ is indicated by the longer grey arrows with shades. The skew scattering probabilities for spin-up and spin-down conduction electrons, resulting in anomalous Hall or inverse spin Hall voltage, are represented by red and blue arrows respectively. } \label{fig1}
\end{center}
\end{figure}

 \

\noindent
\textbf{Results}

\noindent
\textbf{AHE and ISHE of NiPd alloys.}

We measured the SHE of a weakly ferromagnetic NiPd nanowire with 7\%, 8\%, and 9\% Ni using the nonlocal spin injection technique~\cite{14,15,16} at $5 - 50$~K. Figures~2a and 2b show the schematic structure and scanning electron microscopy (SEM) image of the lateral spin valve with the NiPd middle wire. The NiPd alloy is a weak ferromagnet when the Ni concentration is larger than 2.3\%~\cite{17}. Depending on the Ni concentration, its $T_{\rm C}$ varies from a few~K to hundreds of~K. In order to characterize the magnetic property of NiPd wires in our devices, AHE measurements were performed for the reference NiPd Hall bars prepared at the same time as the SHE devices. A representative AHE resistance $R_{xy}$ for Ni$_{0.08}$Pd$_{0.92}$, which is proportional to the magnetization, is shown in Fig.~3a. We note that in the AHE measurements, the external field $H$ was applied perpendicularly to the plane, which is different from the ISHE measurements (see Fig.~2b). At $T = 5$~K a hysteresis loop of the magnetization can be clearly seen. With increasing temperature, the magnetization continually decreases to zero and $R_{xy}$ simply shows the ordinary Hall resistance above 30~K. The Curie temperature $T_{\rm C}$ for Ni$_{0.08}$Pd$_{0.92}$ can be determined to be about 21~K by plotting the remnant magnetization $M_{\rm r}$ and coercive field $H_{\rm c}$ versus temperature (Fig.~3b), which is lower than the value for the bulk alloy~\cite{18}. This is presumably due to the imperfect distribution of Ni in our alloy, prepared by the ion implantation technique, compared with bulk alloys (see Methods). We also performed the same measurements for the other two concentrations where $T_{\rm C}$ for Ni$_{0.07}$Pd$_{0.93}$ and Ni$_{0.09}$Pd$_{0.91}$ are 16 and 32~K, respectively.

\begin{figure}
\begin{center}
\includegraphics[width=6cm]{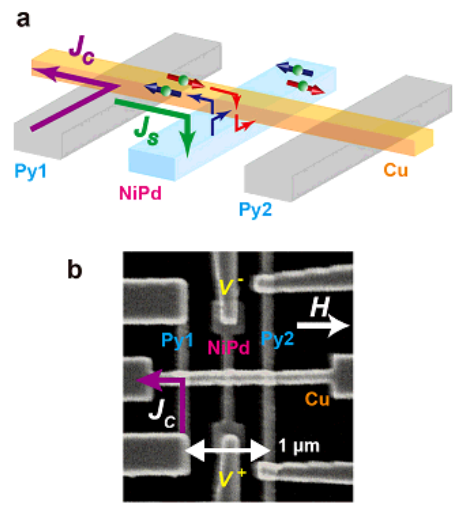}
\caption{\textbf{Structure of the spin Hall device.} \textbf{a}, Schematic image of the device. The green spheres with the red and blue arrows correspond to spin-up and spin-down electrons, respectively. $J_{\rm C}$ and $J_{\rm S}$ are the charge and spin current respectively. The thinner red and blue arrows indicate the spin-up and spin-down electron motion, respectively. \textbf{b}, SEM image of the device consisting of two Py wires and a NiPd middle wire bridged by a Cu wire. The arrow for the magnetic field H represents the positive direction in the case of ISHE measurement.} \label{fig2}
\end{center}
\end{figure}

The ISHE is based on the spin to charge current conversion, which follows the vector product as below: $\boldsymbol{J}_{\rm C} \propto \boldsymbol{J}_{\rm S} \times \boldsymbol{s}_{\rm C}$ where $\boldsymbol{J}_{\rm C}$ and $\boldsymbol{J}_{\rm S}$ are the charge and spin current density, $\boldsymbol{s}_{\rm C}$ is the spin orientation of the conduction electron. Our device configuration is depicted in Fig.~2a. The charge current flows from one of two Ni$_{0.81}$Fe$_{0.19}$ (hereafter Py) wires, i.e., Py1 to the left hand side of the Cu wire. Since the spin-orbit (SO) interaction of Cu is quite weak, it works as a transmitter of the generated spin current at the Cu/Py1 interface to the right hand side of the Cu wire. The spin current is preferentially absorbed into the NiPd middle wire below the Cu wire because of its strong SO interaction. This absorption is confirmed by a reduction of the nonlocal spin valve (NLSV) signal detected at the second Py electrode Py2, as detailed in refs.~15 and 16. In the case of Ni$_{0.08}$Pd$_{0.92}$, the NLSV signal is reduced to 0.25 compared to the value without the NiPd middle wire, and the spin diffusion length is estimated to be 10~nm. The absorbed spin-up and down electrons with the orientation $\textbf{s}_{\rm C}$ parallel to the hard direction of the Py wires are scattered by Ni impurities in Pd, resulting in the Hall voltage $V_{\rm ISHE}$. From the field variation, the overall change of ISHE resistance $R_{\rm ISHE}$ (equal to $V_{\rm ISHE}$ divided by the charge current) can be observed.

\begin{figure*}
\begin{center}
\includegraphics[width=15cm]{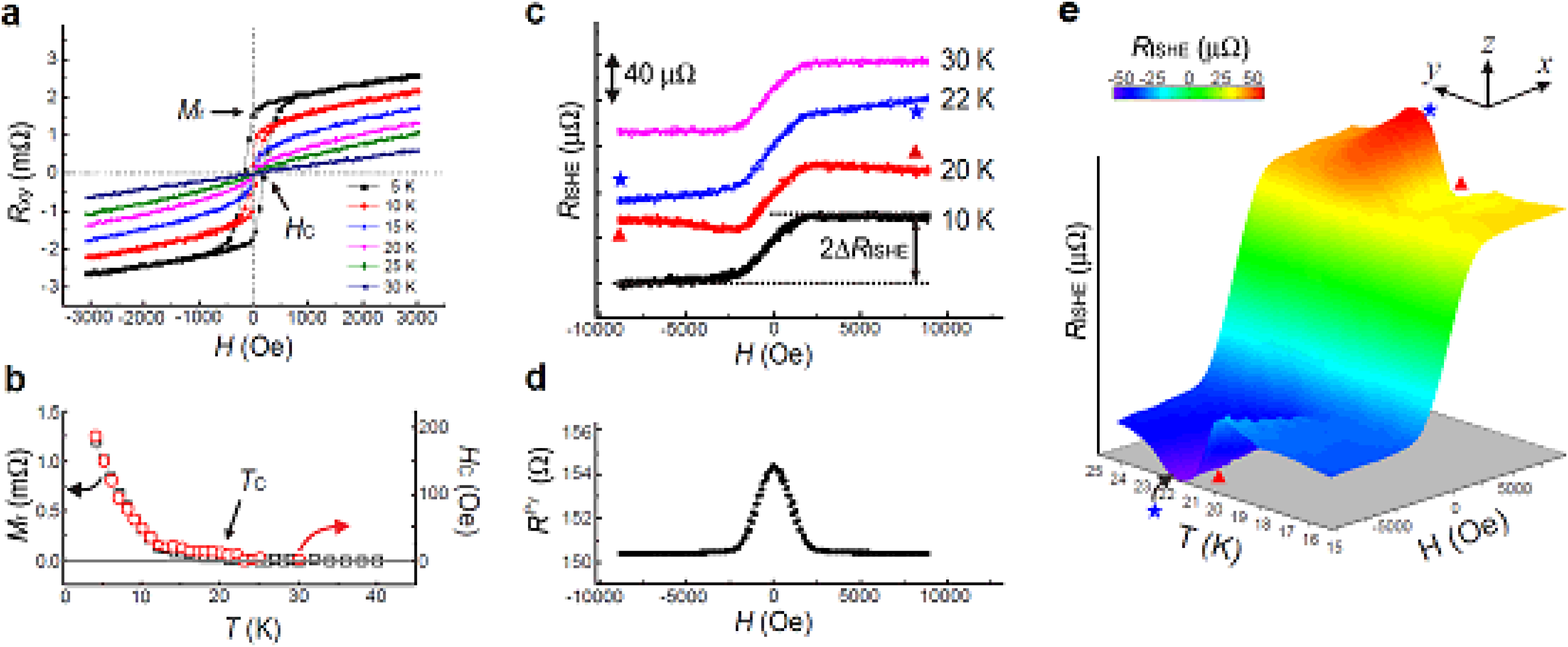}
\caption{\textbf{Temperature dependence of the anomalous Hall and the inverse spin Hall effects of Ni$_{0.08}$Pd$_{0.92}$.} \textbf{a}, AHE resistance $R_{xy}$ at different temperatures. The remnant magnetization $M_{\rm r}$ and coercive field $H_{\rm c}$ correspond to $R_{xy}$ at $H = 0$~Oe and the $x$-intercept of $R_{xy}$ curve, respectively. \textbf{b}, Temperature dependence of $M_{\rm r}$ and $H_{\rm c}$. From both plots $T_{\rm C}$ can be determined to be 21~K. \textbf{c}, Representative RISHE curves of Ni$_{0.08}$Pd$_{0.92}$ measured at 10, 20, 22 and 30~K. At 20 and 22~K, $R_{\rm ISHE}$ at high fields is reduced and enhanced, respectively. \textbf{d}, A typical AMR of Py1 measured at 10~K. \textbf{e}, 3D plot of $R_{\rm ISHE}$ as a function of temperature ($15-25$~K) and magnetic field, repre-sented by both height and color.} \label{fig3}
\end{center}
\end{figure*}

We show representative ISHE resistance $R_{\rm ISHE}$ curves for Ni$_{0.08}$Pd$_{0.92}$ as a function of the magnetic field at various temperatures in Fig.~3c, where all curves are shifted for clarity. Generally $R_{\rm ISHE}$ increases linearly with the magnetic field up to 2000~Oe and then becomes flat. This is consistent with the saturation of the magnetization of Py1, indicated by the anisotropic magnetoresistance (AMR) curve of Py1 shown in Fig.~3d. In the following we focus on $R_{\rm ISHE}$ with saturated magnetization of Py1 (i.e., $|H| > 2000$~Oe). For temperatures far below (10~K; black curve) and above (30~K; pink curve) $T_{\rm C}$, $R_{\rm ISHE}$ is completely flat at high enough fields, which is as expected in the ISHE for nonmagnetic systems. The values of $R_{\rm ISHE}$ are much larger than those for pure Pd, where an intrinsic mechanism should apply~\cite{16,19,20} and we will argue that this increase comes from the extrinsic effect of the Ni impurities. 

For temperatures near $T_{\rm C}$, in contrast, anomalous behavior can be clearly seen. At 20~K just below $T_{\rm C}$ ($\approx 21$~K), $R_{\rm ISHE}$ decreases with increasing the magnetic field ($|H|$) up to 8000~Oe and then flattens out (see the red curve in Fig.~3c). Just above $T_{\rm C}$, at 22~K, on the other hand, $R_{\rm ISHE}$ has a positive slope versus magnetic field even above 2000~Oe and is not yet saturated at our maximum field. By making many similar scans for a full temperature range from 10 to 30~K, we can display this behavior more clearly seen in the 3D plot shown in Fig.~3e. The $x$ and $y$ axes are the magnetic field and temperature, respectively, while both the $z$ axis and color represent $R_{\rm ISHE}$. When the temperature crosses $T_{\rm C}$, a reduction first appears at 20~K as a light blue peak at negative fields or a light yellow dip at positive fields (see red triangles in Fig.~3c); then an enhancement appears at 22~K as a dark blue dip at negative fields or a red peak at positive fields (see blue stars in Fig.~3c). Such a behavior of $R_{\rm ISHE}$ is observed below and above but only in the vicinity of $T_{\rm C}$. As already mentioned, $R_{\rm ISHE}$ for a nonmagnetic metal should be completely flat at any temperature. It is thus unambiguous that the observed change in $R_{\rm ISHE}$ of the NiPd alloys near $T_{\rm C}$ is correlated to the magnetic phase transi-tion of the NiPd wire (see Fig.~3b). On the other hand, such an anomaly near $T_{\rm C}$ is not observed in the temperature dependence of the spin absorption rate or the spin diffusion length of NiPd. This is because the spin diffusion length depends simply on the resistivity and the SO interaction of the NiPd wire while the ISHE signal is strongly affected by the type of scattering mechanisms, i.e. skew scattering, side-jump or band structure. 

 \

\noindent
\textbf{Anomaly in ISHE of NiPd alloys near Curie temperatures.}

In order to better quantify the contribution to the ISHE intrinsic to the NiPd wire, we define the change in $R_{\rm ISHE}$ as $\Delta R_{\rm ISHE} \equiv [R_{\rm ISHE} (H_{\rm max}) - R_{\rm ISHE} (-H_{\rm max})] / 2$, as shown in Fig.~3c. As a maximum field $H_{\rm max}$ we chose 8800~Oe in this work. Certainly $R_{\rm ISHE}$ at 22~K is not fully saturated even at 8800~Oe, which is different from $R_{\rm ISHE}$ at 20~K. This is related to the fact that it is much more difficult to align the localized magnetic moments of Ni impurities above $T_{\rm C}$ compared to those below $T_{\rm C}$ even if high magnetic fields are applied. In Fig.~4a, we plot $\Delta R_{\rm ISHE}$ of Ni$_{0.08}$Pd$_{0.92}$ (shown by red dots) as a function of temperature. The anomaly of $\Delta R_{\rm ISHE}$ near $T_{\rm C}$ is clearly seen. Below 18~K, $\Delta R_{\rm ISHE}$ is almost constant with temperature. Only in the vicinity of $T_{\rm C}$, however, it shows a dip and peak structure at 20 and 21.5~K, respectively. Since $\Delta R_{\rm ISHE}$ varies little with temperature except near $T_{\rm C}$, it can be separated into the temperature-independent background $\Delta R_{\rm ISHE}^{0}$ and the anomalous part $\delta \Delta R_{\rm ISHE}$, as $\Delta R_{\rm ISHE} = \Delta R_{\rm ISHE}^{0} + \delta \Delta R_{\rm ISHE}$. The results for NiPd with 7\% and 9\% of Ni are plotted with black squares and blue triangles, respectively. The increase of $\Delta R_{\rm ISHE}^{0}$ with the doping concentration suggests the extrinsic contribution induced by the Ni impurities. Thus $\Delta R_{\rm ISHE}^{0}$ comes from both the extrinsic SHE of Ni impurities and the intrinsic SHE of Pd~\cite{16}. Despite the fact that $\Delta R_{\rm ISHE}^{0}$ depends strongly on the Ni concentrations, $\delta \Delta R_{\rm ISHE}$ does so primarily via a shift to be near the corresponding $T_{\rm C}$. We emphasize that such critical behavior was reproducible for each of at least three samples at each Ni concentration; thus our arguments are based on more than 10 complete data sets, of which only one is shown in Fig.~3e.

For the three concentrations, $\delta \Delta R_{\rm ISHE}$ are plotted as a function of the reduced temperature $(T-T_{\rm C})/T_{\rm C}$ in Fig.~4b. Note that each curve corresponds to one full data set: for example the red line represents a reduction of all of Fig.~3e. Although the three devices have different $T_{\rm C}$ and $\Delta R_{\rm ISHE}^{0}$, the anomalous behavior appears to be universal near $T_{\rm C}$: the three curves almost scale onto one. The amplitudes of the dip and peak are about 5~$\mu \Omega$. The anomalous behavior disappears when the reduced temperature differs from $T_{\rm C}$ by 10\%. These results are obviously different from the case of the AHE~\cite{13} where the anomaly of the Hall resistivity $\rho_{\rm H}$ appears only below $T_{\rm C}$. 

\begin{figure}
\begin{center}
\includegraphics[width=7cm]{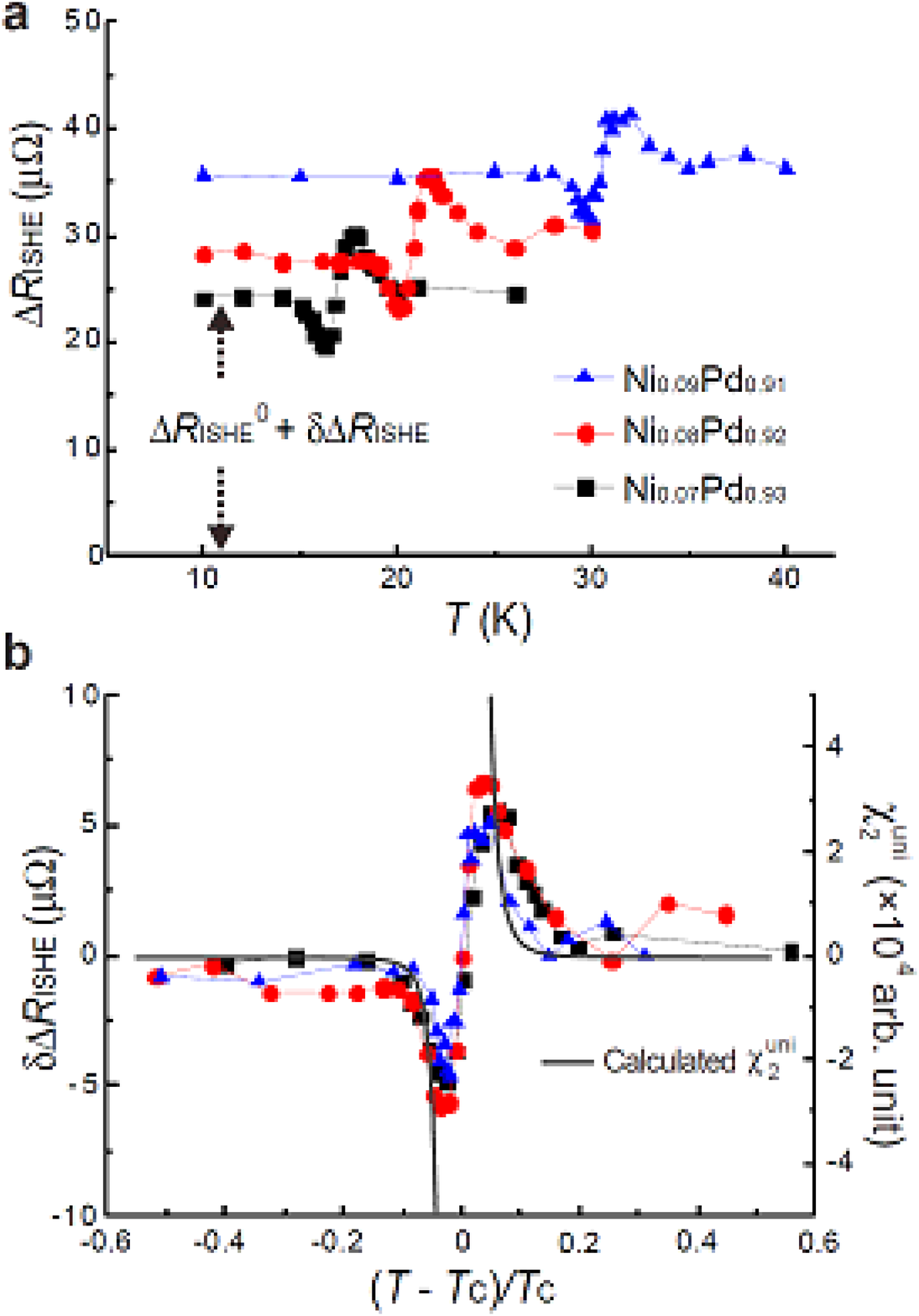}
\caption{\textbf{The anomaly in the inverse spin Hall effect near $T_{\rm C}$.} \textbf{a}, Inverse spin Hall resistance $\Delta R_{\rm ISHE}$ of Ni$_{x}$Pd$_{1-x}$ middle wires with Ni concentration $x = 0.07$ (black square), 0.08 (red dot) and 0.09 (blue triangle) as a function of temperature. The anomaly of $\Delta R_{\rm ISHE}$ for each Ni concentration appears only near its $T_{\rm C}$. \textbf{b}, The anomalous component $\delta \Delta R_{\rm ISHE}$ for Ni concentration equal to $x = 0.07$ (black square), 0.08 (red dot) and 0.09 (blue triangle) plotted as a function of normalized temperature $(T - T_{\rm C})/T_{\rm C}$. For comparison we also show the uniform second order nonlinear susceptibility $\chi_{2}^{\rm uni}$ (black solid line).} \label{fig4}
\end{center}
\end{figure}

 \

\noindent
\textbf{Discussion}

What is the origin of the anomaly of $\delta \Delta R_{\rm ISHE}$ observed only in the vicinity of $T_{\rm C}$? Resistive anomalies near magnetic critical points $T_{\rm C}$ have been studied for a long time. The anomaly near $T_{\rm C}$ in the longitudinal resistivity appears within the first Born approximation from quadratic fluctuations $\langle (M_{i} - \langle M_{i} \rangle)(M_{j} -\langle M_{j} \rangle)\rangle$ in the moments $M_{i}$ and $M_{j}$ of localized spins at sites $i$ and $j$ where contributions come only when $i$ and $j$ are within a certain cut-off distance~\cite{21}. The resistivity then depends on a linear local magnetic susceptibility $\chi_{0}^{\rm loc}$ which is a summation of short-range fluctuations up to this distance. To simplify notation we shall denote the sum of such short range correlations as $\sum \langle (M -\langle M \rangle)^{2}\rangle$, omitting the site subscripts, and we shall use similarly abbreviated notations for higher order moments. As mentioned in the introduction, the anomaly of $\rho_{\rm H}$ in the AHE near $T_{\rm C}$ was explained by Kondo~\cite{13}. Note that the quantities $r_{1}$ and $r_{2}$ in ref.~13 correspond to a first order nonlinear susceptibility $\chi_{1}^{\rm loc} \sim \sum \langle (M -\langle M \rangle)^{3}\rangle$ and a second order nonlinear susceptibility $\chi_{2}^{\rm loc} \sum \langle (M -\langle M \rangle)^{4}\rangle$, respectively. We change notation for simplicity, and also to emphasize that Kondo makes a strictly local approximation, whereas we extend the theory to include nearby correlations between atomic moments. The anomaly of $\rho_{\rm H}$ in the AHE appears from the first order fluctuation of local moments $\chi_{1}^{\rm loc}$. The second order spin fluctuation $\chi_{2}^{\rm loc}$ is also defined in ref.~13 but made no contribution to the AHE. This will be explained below; we first note that the $\chi_{1}^{\rm loc}$ and $\chi_{2}^{\rm loc}$ terms appear in the $s$-$d$ Hamiltonian to the same order with respect to the spin-orbit coupling constant $\lambda$ of the localized moment. The term in $\chi_{1}^{\rm loc}$ changes sign for spin-up or down electrons, while $\chi_{2}^{\rm loc}$ does not. Thus the scattering amplitudes for spin-up and down conduction electrons are different and proportional to $\chi_{1}^{\rm loc} + \chi_{2}^{\rm loc}$ and $-\chi_{1}^{\rm loc} + \chi_{2}^{\rm loc}$ respectively~\cite{13}. As illustrated in Fig.~1, in the case of the AHE configuration, spin-up and down conduction electrons are scattered to opposite directions. As a result, $\rho_{\rm H}$ is proportional to the difference of scattering amplitudes between spin-up and down electrons, i.e., $\chi_{1}^{\rm loc}$. On the other hand, in the ISHE configuration, both spin-up and down electrons are scattered to the same side. Unlike the AHE, the obtained $R_{\rm ISHE}$ is proportional to the sum of scattering amplitudes between spin-up and down electrons, i.e., $\chi_{2}^{\rm loc}$. The effect of such terms has been hidden for 50 years~\cite{13}. However, it is naturally expected that the contributions related to $\chi_{2}^{\rm loc}$ can survive if the incident current is the spin current, which is exactly the case for the ISHE. 

To make a comparison between theory and experiment we should, in principle, calculate $\chi_{2}^{\rm loc}$, which is integrated over the Fermi surface. This $\chi_{2}^{\rm loc}$ should be proportional to the experimental $\delta \Delta R_{\rm ISHE}$. However, the integration depends essentially on the Fermi wave vector, which acts as a cut-off to any divergence. To avoid including material dependent parameters, we compare to a quantity, the uniform second order nonlinear susceptibility $\chi_{2}^{\rm uni}$ which is independent of cut-off. $\chi_{2}^{\rm uni}$ is the response of the moments on all the sites to a uniform magnetic field, and is the sum of correlations over all distances without any cut-off (see Supplementary Discussion). In Fig.~4b we superimpose on the experimental data a mean field calculation of $\chi_{2}^{\rm uni}$. It diverges with opposites signs above and below $T_{\rm C}$. We interpret the dip and peak observed in the experimental data as a smearing of such a divergence. In a full calculation the theoretical spin Hall resistance should also be smeared: the divergence is cut-off, just as for the longitudinal resistance~\cite{21}. To confirm the above scenario, we have performed Monte Carlo simulations for $\chi_{2}^{\rm loc}$ where the contribution of nearest-neighbor correlations is taken into account. As can be seen in Supplementary Figure~S1, the smearing of the divergence of $\chi_{2}^{\rm loc}$ at $T_{\rm C}$ is well-reproduced. Thus we argue that neither the full theory, nor the experiments should diverge at $T_{\rm C}$. 

Finally we note that the anomaly of $\rho_{H}$ in the AHE in pure Fe~\cite{11} and Ni~\cite{11,12} may alternatively be explained by Karplus and Luttinger's theory which is based on band effects~\cite{22}. In this theory $\rho_{\rm H}$ would be proportional to $\rho_{xx}^{2} \langle M \rangle$ where $\rho_{xx}$ is the longitudinal resistivity and $\langle M \rangle$ is the averaged magnetization. Since $\langle M \rangle$ is zero above $T_{\rm C}$, the sign change of $R_{\rm ISHE}$ below and above $T_{\rm C}$ seems to preclude explanation by such band effects. 

In conclusion, we have demonstrated a characteristic behavior of the spin Hall effect at the magnetic phase transition of the weak ferromagnetic NiPd near its Curie temperature $T_{\rm C}$. It can be explained by the fluctuation contributions to skew scattering via spin-orbit interactions, and appears as the second order nonlinear spin fluctuation $\chi_{2}^{\rm loc}$. Considering the small volume of NiPd wire ($10^{-16}$~cm$^{3}$) and the magnetic moment of NiPd, the total magnetic moment involved in the experiment is extremely small: less than $10^{-14}$~emu. Such a tiny moment is far below the sensitivity of magnetometer (for instance, $10^{-8}$~emu for SQUID). The successful extractions of the contribution of higher order spin fluctuations in our experiments prove the extremely high sensitivity of our method. This technique could be very useful to understand more complex spin textures such as spin-glasses and spin-ice. For those systems, the linear magnetic susceptibility $\chi_{0}^{\rm uni}$ has been well-studied so far but there have been almost no experimental studies of the higher order spin fluctuations where some important physics might be still hidden. The present method constitutes a novel approach to detect high order nonlinear fluctuations with an extremely high sensitivity, which is impossible to achieve by a direct measurement. This will broaden the application of the SHE from potential devices to a means of measuring higher order spin fluctuations, particularly in the critical region.

 \

\noindent
\textbf{Method}

\noindent
\textbf{Sample preparation.}

The spin Hall devices were fabricated on a thermally oxidized silicon substrate using electron beam lithography on poly-methyl-methacrylate (PMMA) resist and a subsequent lift-off process. Here we used a lateral spin valve structure which consists of two Py wires (30~nm thick and 100~nm wide) with a separation distance of 1~$\mu$m and a NiPd middle wire (20~nm thick and 100~nm wide) bridged by a Cu wire (100~nm thick and 100~nm wide) as is detailed in Fig.~2 in the main text. A Pd wire was first deposited by electron beam evaporation using a 99.999\% purity source. In order to form the NiPd alloy wire with a Ni content of 7\%, 8\% and 9\%, we implanted a suitable amount of Ni$^{+}$ into the Pd wire using an Ion Implanter (NH-20SR) with an acceleration voltage of 30~kV. As is mentioned in the main text, the prepared NiPd alloy is not perfectly homogenous and nickel is distributed over $\sim 10$~nm below the Cu/NiPd interface in 20~nm thick Pd wires. This nickel distribution would lower the $T_{\rm C}$ compared to bulk NiPd alloys. However, the spin diffusion length of NiPd of about 10~nm corresponds to the maximum distribution of Ni in Pd. That is why we could still observe the spin fluctuation of Ni moment. The two Py wires were deposited by electron beam evaporation, while the Cu bridge was fabricated by a Joule heating evaporator. Prior to the Cu evaporation, a careful Ar ion beam milling with a beam current of 12~mA and beam voltage of 600~V was performed for 30~seconds to clean the surfaces of Py and NiPd wires and to obtain highly transparent contacts.

 \

\noindent
\textbf{Acknowledgements}

We would like to thank Y. Iye and S. Katsumoto for the use of the lithography facili-ties. T.Z. also thanks the Advanced Science Research Center of the JAEA for support via the REIMEI program. This work was supported by KAKENHI and a Grant-in-Aid for Scientific Research in Priority Area from MEXT.

\noindent
\textbf{Author Contributions}

D. W. and Y. N. designed the experiments and fabricated devices. D. W. collected the data and carried out the analysis. B. G., T. Z., and S. M. developed the theoretical analysis. B. G. developed the detailed equations in Supplementary Information. Y. O. planned and supervised the project. All authors discussed the results and commented on the manuscript.

\end{document}